\documentclass{appolb}

\usepackage{graphicx}

\renewcommand{\Re}{\mathop{\mathrm{Re}}}
\renewcommand{\Im}{\mathop{\mathrm{Im}}}

\begin{document}
\title{Charmonium spectral functions in $\bar p A$ collision%
\thanks{Presented at Excited QCD 2017, Sintra, Portugal}%
}
\author{Gy. Wolf, G. Balassa, P. Kov\'acs, M. Z\'et\'enyi,
\address{\vspace*{-0.4cm} Wigner RCP, Budapest, 1525 POB 49, Hungary}
\\[0.6cm]
Su Houng Lee
\address{\vspace*{-0.4cm} Department of Physics and Institute of Physics and Applied Physics, Yonsei University, Seoul 120-749, Korea}
}
\maketitle
\begin{abstract}
  We study the in-medium propagation of low-lying charmonium states:
  J/$\Psi$, $\Psi$(3686), and $\Psi$(3770) in a $\bar p$ Au 10 GeV
  collision. This energy regime will be available for the PANDA
  experiment.  The time evolution of the spectral functions of the
  charmonium states is studied with a BUU type transport model.  We
  observe a substantial effect of the medium in the dilepton spectrum.
\end{abstract}
  
\section{Introduction}

The observation of charmonium in vacuum and in medium is an important goal of
the PANDA collaboration at the future FAIR complex. In antiproton induced
reactions large number of charmed states are expected to be created.
Furthermore, $\bar p A$ reactions are best suited to observe charmed particles
in nuclear matter, since the matter is in this case much simpler than the one
created in a heavy ion collision. 

The spectral functions of the J/$\Psi$, $\Psi$(3686), and $\Psi$(3770) vector
mesons are expected to be modified in a strongly interacting environment.
In our transport model of the BUU type the time evolution of 
single-particle distribution functions of various hadrons are evaluated within
the framework of a kinetic theory. The $\Psi$(3770) meson is
already a broad resonance in vacuum, while the J/$\Psi$, $\Psi$(3686) mesons
may acquire a noticeable
width in nuclear matter \cite{SHLee_Charm}. Therefore, one
has to propagate properly the spectral functions of these charmonium states.
This is the main goal of our paper. Similar investigation has been carried out
in \cite{Golubeva_Charm}

\section{Off-shell transport of broad resonances}

If we create a particle in a medium with in-medium mass, through the evolution,
it should regain its vacuum mass, when it reaches vacuum. If we use a local
density approximation for changing its mass, we clearly break the energy
conservation. We need for the propagation of off-shell particles a more
sophisticated method.
We can describe the in-medium properties of particles with
an ``off-shell transport''. These equations are derived by starting from
the Kadanoff-Baym equations for the Green's functions of particles. 
Applying first-order gradient expansion after a Wigner transformation 
\cite{Cassing-Juchem00,Leupold00} one arrives at a transport equation for 
the retarded Green's function.  
In the medium, particles acquire a self-energy $\Sigma(x,p)$  which depends on position 
and momentum  as well as the local properties of the surrounding medium.
Particle properties are described by their spectral function 
being the imaginary part of the retarded propagator
\begin{equation}
    \label{spectral}
    {\cal A }(p) = -2 \Im G^{ret}(x,p) = 
      \frac{\hat{\Gamma}(x,p)}{(E^2 -{\vec p}^2-m_0^2-
          \Re\Sigma^{ret}(x,p))^2
       + \frac{1}{4}\hat{\Gamma}(x,p)^2}\,,
\end{equation}
where the resonance widths $\Gamma$ and $\hat{\Gamma}$ are related 
via  $\hat{\Gamma}(x,p) = - 2 \Im \Sigma^{ret}\approx 2m_0\Gamma$,
and $m_0$ is the vacuum pole mass of the respective particle. 

To solve numerically the Kadanoff-Baym equations one may exploit the 
test-particle ansatz  for the retarded Green's function  
\cite{Cassing-Juchem00,Leupold00}. This function can be interpreted as
a product of particle number density multiplied with the spectral function 
${\cal A}$.

The relativistic version of the equation of motion have been derived
in \cite{Cassing-Juchem00}:
\begin{eqnarray}
  \label{eq:x}
  \frac{d {\vec x}}{dt} & = &
  \frac{1}{1 - C}
  \frac{1}{2 E}
  \left(
    2 {\vec p} + {\vec \partial}_{p} \Re \Sigma^{ret} +
    \frac{m^2 - m_0^2 - \Re \Sigma^{ret}}{\hat{\Gamma}}
    {\vec \partial}_{p} \hat{\Gamma}
  \right),\\
  \label{eq:p}
  \frac{d {\vec p}}{d t} & = &
  - \frac{1}{1-C}
  \frac{1}{2 E}
  \left(
    {\vec \partial}_{x} \Re \Sigma^{ret}
    + \frac{m^2 - m_0^{2}
      - \Re \Sigma^{ret}}{\hat{\Gamma}}
    {\vec \partial}_{x} \hat{\Gamma}
  \right),\\
  \label{eq:e}
  \frac{d E}{d t} & = &
  \frac{1}{1-C}
  \frac{1}{2 E}
  \left(
    \partial_{t} \Re \Sigma^{ret}
    + \frac{m^2 - m_0^{2}
      - \Re \Sigma^{ret}}{\hat{\Gamma}}
    {\partial}_{t} \hat{\Gamma}
  \right),
\end{eqnarray}
with the renormalization factor
\begin{equation}
  \label{eq:C}
  C = \frac{1}{2 E}
  \left(
  {\partial_E}\Re \Sigma^{ret}
  + \frac{m_n^2 - m_0^2 - \Re \Sigma^{ret}}{\hat{\Gamma}}
  {\partial_E}\hat{\Gamma}
\right).
\end{equation}
Above, $m = \sqrt{E^2 - {\vec p}^2}$ is the mass of an individual
test-particle.  The $\Sigma^{ret}$ self-energy is considered to be a
function of the $n$ baryon density, the $E$ energy, and the $\vec p$
momentum.

Change of the $m$ test-particle mass can be more clearly seen by
combining Eqs.~(\ref{eq:p}) and (\ref{eq:e}) to result in
\begin{equation}
   \label{eq:m}
  \frac{dm^2}{dt} \,=\, \frac{1}{1-C} \left(\frac{d}{dt} {\Re \Sigma^{ret}} +
   \frac{m^2 - m_0^2 - \Re \Sigma^{ret}}{\hat \Gamma}
   {\frac{d}{dt} {\hat \Gamma}} \right),
\end{equation}
with the comoving derivative $d/dt \equiv \partial_t + {\vec p}/E
{\vec \partial}_x$.
The vacuum spectral function is recovered when the particle leaves the medium
\cite{Almasi}.

The equations of motion of the test-particles 
have to be supplemented by a collision term
which couples the equations of the different particle species.
It can be shown \cite{Leupold00} that 
the collision term has the same form as in the
standard BUU treatment.

In our calculations we  employ a simple 
form of the self-energy of a vector meson $V$:
\begin{eqnarray} \label{areal}
{\rm \Re} \Sigma^{ret}_V & = & 2 m_V \Delta m_V \frac{n}{n_0},\\
\label{aimag}
{\rm \Im} \Sigma^{ret}_V & = & -m_V (\Gamma^{vac}_V + \Delta \Gamma_V \frac{n}{n_0}).
\end{eqnarray}
Eq.~(\ref{areal}) describes a ''mass shift'' 
$\Delta m = \sqrt{m_V^2+\Re\Sigma_V^{ret}}-m_V\approx  \Delta m_V \frac{n}{n_0}$.
The imaginary part contains the vacuum width $\Gamma^{vac}_V$.
The second term in Eq.~(\ref{aimag}) results from the collision broadening. 
The parameters ($\Delta m_V, \Delta \Gamma_V)$ are taken from \cite{SHLee_Charm} and
are the following,
\begin{table}[th]
\centering
\begin{tabular}
[c]{|c|c|c|}\hline
Charmonium &  $\Delta m_V$ & $\Delta \Gamma_V$\\\hline
J/$\Psi$ & -15 MeV & 20 MeV \\\hline
$\Psi$(3686)& -100 MeV& 20 MeV \\\hline
$\Psi$(3770) & -140 MeV & 20 MeV \\\hline
\end{tabular}
\label{tab:med}
\end{table}

If a meson is generated at normal density its mass is distributed in
accordance with the in-medium spectral function. If the meson
propagates into a region of higher density then the mass will be
lowered according to the action of $\Re\Sigma^{ret}$ in
Eq.~(\ref{eq:m}). However if the meson comes near the threshold the
width $\hat{\Gamma}$ becomes very small and the second term of the
right hand side of Eq.~(\ref{eq:m}) dominates, so reverses this trend
leading to an increase of the mass. This method is energy conserving.
We note that the propagation of $\omega$ and $\rho$ mesons at the
HADES energy range have been investigated in \cite{KampferWolf-2010}
with the same method.

\section{Results}
We applied the Eqs.~(\ref{eq:x},\ref{eq:p},\ref{eq:e}) to propagate
the test particles for charmonium states using our BUU code
\cite{KampferWolf-2010,Wolf90,Wolf93}. The cross section for
charmonium production in the different channels have no relevance in
this study, since we do not calculate the background and we will not
add absolute normalization.

\begin{figure}[htb]
\centerline{%
  \includegraphics[width=0.45\textwidth]{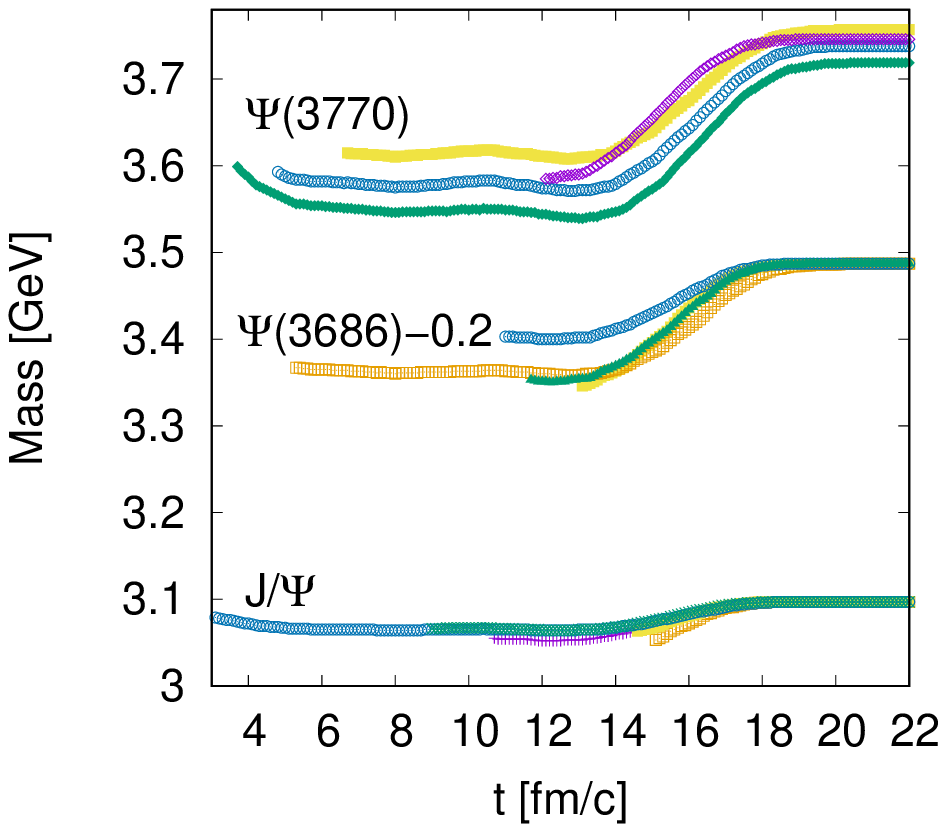}
  \includegraphics[width=0.45\textwidth]{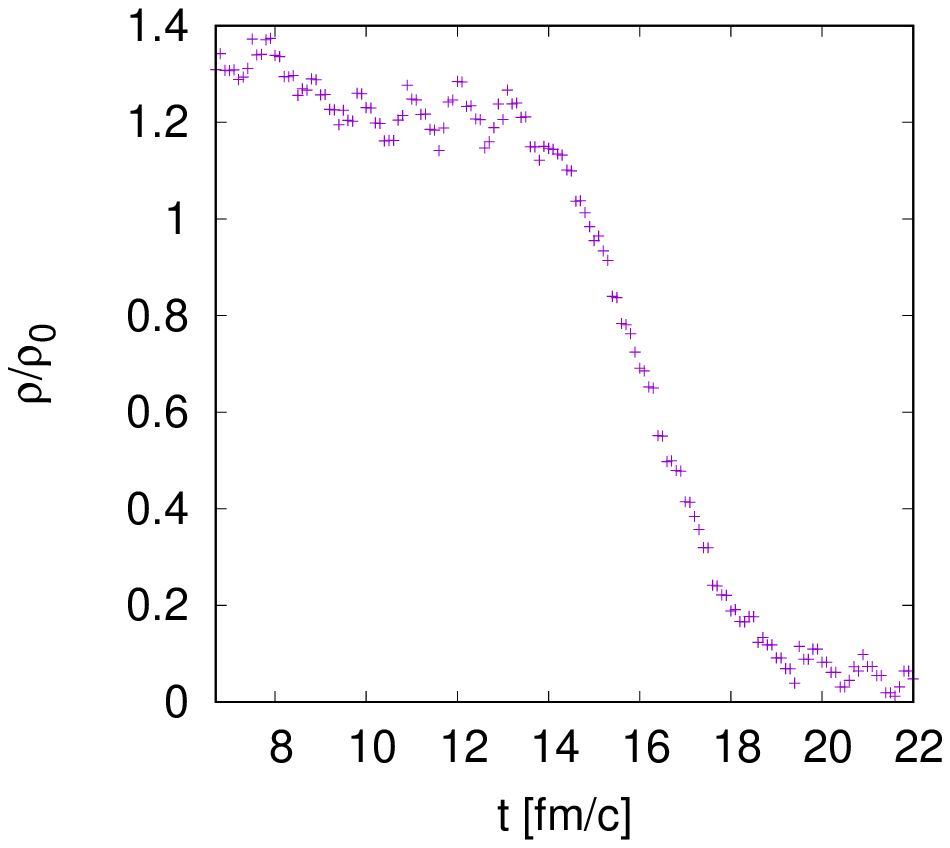}
}
\caption{In the left panel the evolution of the masses of J/$\Psi$,
  $\Psi$(3686), $\Psi$(3770) is shown.  For the better visibility the
  masses of the $\Psi$(3686) are shifted downwards artificially by 200
  MeV. In the right panel we show the average density as a function of
  time felt by the charmonium states.}
\label{Fig:Charmonium_evol}
\end{figure}

In the left panel of Fig.~\ref{Fig:Charmonium_evol} we show how the
masses of test particles representing charmonium mesons are developing
in $\bar p A$ collisions. Note that for getting a better overview in
the figure we have shifted the mass of $\Psi$(3686) downwards by 200
MeV. At the end of the collision process, where the density is very
low, the masses reach the vacuum value as it should be.  The mass of
$\Psi$(3770) spreads even in the end of the collision due to the
substantial vacuum width.  Most of the time the masses of these mesons
are either shifted downwards or at their vacuum value.  The transition
period between them is rather short due to the large charmonium
velocity and the relatively narrow surface.  The evolution of average
density felt by the charmonium states are shown in the right panel of
Fig.~\ref{Fig:Charmonium_evol}. This shows the same thing: the
transition from the dense state to the vacuum one is approximately 4
fm/c.

\begin{figure}[htb]
\centerline{%
  \includegraphics[width=0.7\textwidth]{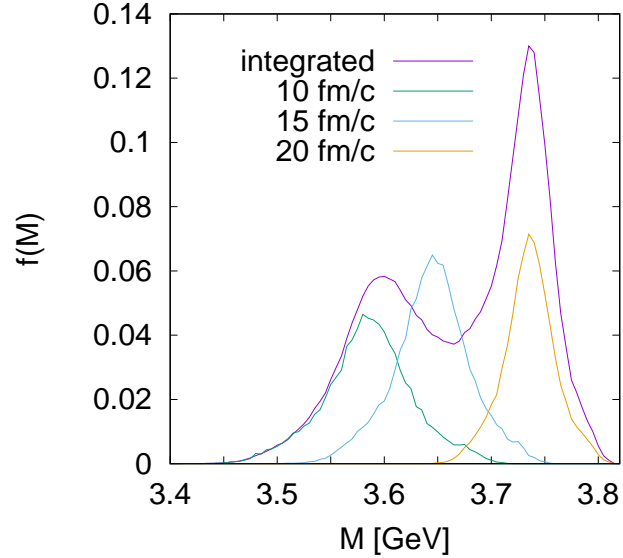}
}
\caption{We show the mass spectra of $\Psi$(3770) at 10, 15 and 20
  fm/c times, and also show the time integrated mass spectra
  normalized arbitrarily.}
\label{Fig:Massspectra}
\end{figure}

In Fig.~\ref{Fig:Massspectra} we show the mass spectra of $\Psi$(3770)
at different times and also the time-integrated mass spectrum. At each
time the mass spectrum can very well be described by a
Breit\,-\,Wigner form. However, the time-integrated spectrum has two
peaks. One corresponds to the vacuum value, the other one to the dense
phase. The valley between them is due to the short transition time.

\begin{figure}[htb]
\centerline{%
  \includegraphics[width=0.7\textwidth]{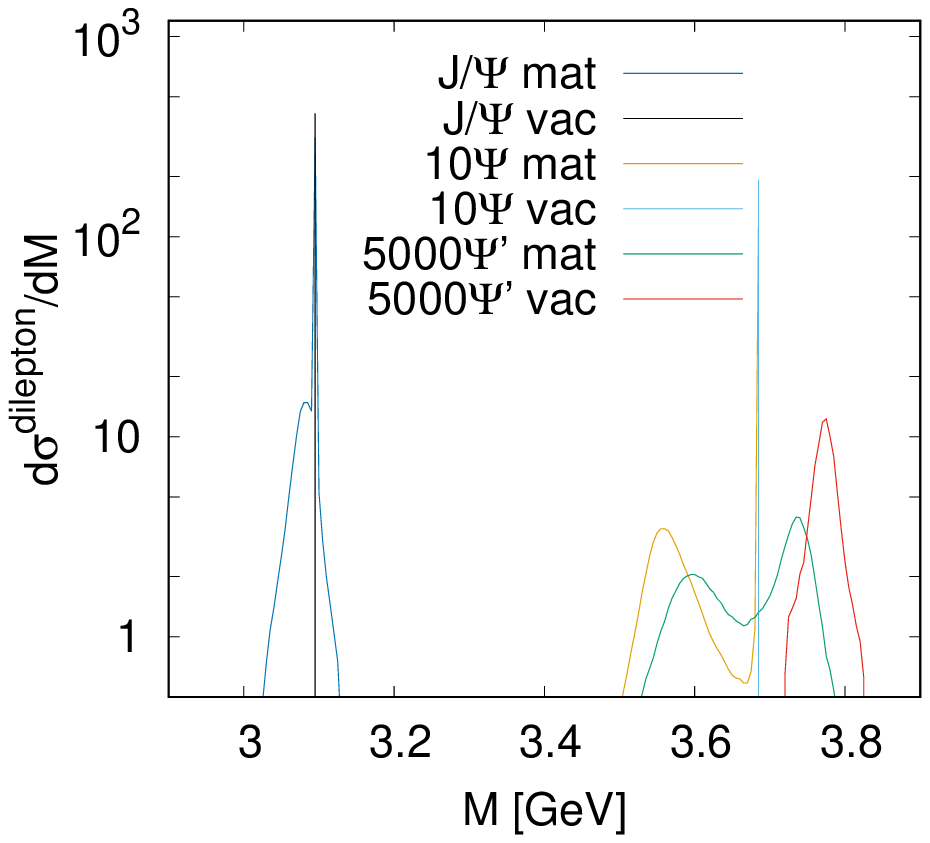}
}
\caption{Charmonium contribution to the dilepton spectra taking into
  account the in-medium modifications and compared it with
  calculations using only vacuum properties. The yields of
  $\Psi$(3770) are multiplied by 5000, and the yields of $\Psi$(3686)
  are multiplied by 10.}
\label{Fig:Dilepspectra}
\end{figure}
In Fig.~\ref{Fig:Dilepspectra} we show the charmonium contributions to
the dilepton spectrum in a $\bar p$ Au 10 GeV collision. We compare
the in-medium and the vacuum calculations. Since we did not introduce
detector resolution, the vacuum calculations for J/$\Psi$ and for
$\Psi$(3686) result in discrete lines. In the in-medium calculations we
can see the two-peak structures for each mesons, however, for the
J/$\Psi$ the effect is negligible, since its mass shift is very
small. For the other two states both peak should be seen in an
experiment. In \cite{Golubeva_Charm} this effect is not there since
they do not have a mass shift for the charmonium states.

\section{Summary}

We calculated the charmonium contribution to the dilepton spectra. We
have shown that via their dileptonic decay there is a good chance to
observe the in-medium modification of the higher charmonium states:
$\Psi$(3686), $\Psi$(3770) in a $\bar p$ Au 10 GeV collision. However,
we have to investigate whether the background (Drell\,-\,Yan process and
the open charm decay) in the dilepton yield allow us to see this
effect.  This energy regime will be available by the PANDA experiment.

\section*{Acknowledgments}

Gy.~W., M.~Z., G.~B., and P.~K. were supported by the Hungarian OTKA
fund K109462 and Gy.~W., M.~Z. and P.~K. by the HIC for FAIR Guest
Funds of the Goethe University Frankfurt.


\end{document}